\RequirePackage{fix-cm}

\documentclass[onecolumn]{svjour3} 
\smartqed  
\usepackage{graphicx}
\usepackage{amssymb} 

\usepackage{subfigure}
\usepackage{amsmath}
\usepackage{amsfonts}
\usepackage{amsbsy}
\usepackage{times}
\usepackage[left]{lineno}

 \journalname{J. Braz. Soc. Mech. Sci. Eng.}
\begin{document}

\title{THE FEEDBACK EFFECT CAUSED BY BED LOAD ON A TURBULENT LIQUID FLOW \thanks{Accepted Manuscript for the Journal of the Brazilian Society of Mechanical Sciences and Engineering, v. 36, p. 725-736, 2014. The final publication is available at Springer via http://dx.doi.org/10.1007/s40430-013-0122-y}}


\author{Erick M. Franklin         \and
        Fab\'iola T. Figueiredo        \and
				Eug\^enio S. Rosa
}


\institute{Erick de Moraes Franklin \at
              Faculty of Mechanical Engineering, University of Campinas - UNICAMP \\
              Tel.: +55-19-35213375\\
              \email{franklin@fem.unicamp.br}           
           \and
           Fab\'iola Tocchini de Figueiredo \at
              Faculty of Mechanical Engineering, University of Campinas - UNICAMP \\
              \email{fabiolafig@fem.unicamp.br}           
					\and
						Eug\^enio Span\'o Rosa \at
              Faculty of Mechanical Engineering, University of Campinas - UNICAMP \\
              \email{erosa@fem.unicamp.br}           
}

\date{Received: date / Accepted: date}

\maketitle

\begin{abstract}
Experiments on the effects due solely to a mobile granular layer on a liquid flow are presented (feedback effect). Nonintrusive measurements were performed in a closed conduit channel of rectangular cross section where grains were transported as bed load by a turbulent water flow. The water velocity profiles were measured over fixed and mobile granular beds of same granulometry by Particle Image Velocimetry. The spatial resolution of the measurements allowed the experimental quantification of the feedback effect. The present findings are of importance for predicting the bed-load transport rate and the pressure drop in activities related to the conveyance of grains.
\keywords{Turbulent boundary layer \and feedback effect \and sediment transport \and bed load}
\end{abstract}

\section{List of Symbols}
\label{section:nomenclature}

$A$ = constant;\\
$B$ = constant;\\
$C$ = constant; \\
$d$ = grain diameter ($\mu m$); \\
$d_{50}$ = diameter of which $50\%$ of the grains are smaller ($\mu m$); \\
$D_{piv}$ = distance from the bed from which the employed PIV device was able to measure (m); \\
$f$ = friction factor; \\
$g$ = gravity acceleration ($m/s^2$); \\
$H_{gap}$ = distance from the granular bed to the top wall (m); \\
$Q$ = water flow rate ($m^3/h$); \\
$Re$ = channel Reynolds number, $Re=\rho \overline{U}2H_{gap}/\mu$; \\
$Re_*$ = particle Reynolds number, $Re_*=\rho u_*d/\mu$; \\
$u$ = mean velocity of the fluid (m/s); \\
$\overline{U}$ = cross-section mean velocity of the fluid (m/s); \\
$U_{max}$ = maximum velocity of the water profile (m/s); \\
$u_*$ = shear velocity (m/s); \\
$u^+$ = dimensionless velocity, $u^+\,=\,u/u_*$; \\
$W_b$ = bed-load transport rate (kg/s); \\
$y$ = vertical distance (m); \\
$y_0$ = roughness lenght (m); \\
$y^+$ = dimensionless vertical distance, $y^+\,=\,yu_*/\nu$; \\
$y_0^+$ = dimensionless roughness lenght, $y_0^+\,=\,y_0u_*/\nu$; \\

\noindent \textbf{Greek symbols}\\
$\kappa$ = von K\'arm\'an constant; \\
$\nu$ = kinematic viscosity ($m^2/s$); \\
$\rho$ = specific mass of the fluid ($kg/m^3$); \\
$\rho_{s}$ = specific mass of the grains ($kg/m^3$);\\
$\tau$ = shear stress on the bed ($N/m^2$);      and\\
$\theta$ = Shields number, $\theta =\tau / \left( (\rho_{s}-\rho)gd\right)$.

\section{Introduction}

When the shear stresses of a turbulent liquid flow are bounded to some limits, granular matter is entrained by the flow as a mobile layer of grains, known as bed load. The thickness of this mobile layer, whose grains stay in contact with the fixed part of the granular bed, is of a few grain diameters \cite{Bagnold_1,Raudkivi_1,Yalin_1}. This situation is frequently found in nature and in industry. In nature, it is present in the erosion of river banks and in the migration of river bed forms, for example. In industry, the examples are mostly related to closed-conduit flows: the transport of sand in hydrocarbon pipelines and in dredging activities, but also the transport of grains in sewer systems. In these cases, the correct predictions of the bed-load transport rate and of the pressure drop are of importance.

In principle, two dimensionless groups are necessary to determine the bed-load transport rate, although many different bed-load formulas exist and a real consensus about this issue is absent \cite{Gomez_1,Nakato_1,Martin_1}. One dimensionless group frequently employed is the Shields number $\theta$, defined as the ratio of the entraining to the resisting forces

\begin{linenomath*}
\begin{equation}
\theta = \frac{\tau}{(\rho_{s}-\rho)gd}
\label{shields}
\end{equation}
\end{linenomath*}

\noindent where $\tau$ is the shear stress caused by the fluid on the granular bed, $d$ is the grain diameter, $g$ is the gravitational acceleration, $\rho$ is the specific mass of the fluid and $\rho_{s}$ is the specific mass of the grain. In the case of two-dimensional turbulent boundary layers, the shear stress is $\tau\,=\,\rho u_*^2$ , where $u_*$ is the shear velocity, and the near-wall liquid velocity profile holds true \cite{Schlichting_1}. Near the bed,

\begin{linenomath*}
\begin{equation}
u^+\,=\,\frac{1}{\kappa} ln \left( \frac{y}{y_0} \right) \,=\,\frac{1}{\kappa} ln(y^+) +B
\label{mean_velocity}
\end{equation}
\end{linenomath*}

\noindent where $\kappa\,=\,0.41$ is the von K\'arm\'an constant, $y_0$ is the roughness length, $u^+\,=\,u/u_*$ is a dimensionless velocity, $u$ is the mean velocity of the fluid, $y^+\,=\,yu_*/\nu$ is the transversal distance normalized by the viscous length, $\nu$ is the kinematic viscosity and $B$ is a constant. The wake function was not considered in Eq. \ref{mean_velocity} \cite{Tennekes_Lumley}. Although the second and third terms of Eq. \ref{mean_velocity} are equivalent, the second is generally employed for hydraulic rough regimes, while the third is employed for hydraulic smooth regimes. In the latter case, $B=5.5$ \cite{Schlichting_1}.

The second dimensionless group commonly employed is the particle Reynolds number $Re_*$, which is the Reynolds number of the fluid flow at the grain scale

\begin{linenomath*}
\begin{equation}
Re_*\,=\,\frac{u_*d}{\nu}
\label{reynolds}
\end{equation}
\end{linenomath*}

\noindent where $u_*$ appears here as the characteristic velocity of the fluid at the grain scale (it is the fluid velocity at $y\,\approx\,1.5y_0$, according to Eq. \ref{mean_velocity}). In case of liquids, the mobile layer is located in a region close to the fixed part of the bed, that may correspond to the viscous, or to the buffer, or even to the logarithmic sublayers of the turbulent boundary layer. The ratio between the length-scales of the roughness elements $d$ and that of viscosity $\nu /u_*$ determines which sublayer the bed load takes place. This ratio is equal to $Re_*$, so that it is a pertinent parameter concerning the fluid flow near the granular bed.

However, there is a major difficulty concerning the determination of $\theta$ and $Re_*$: the fluid flow depends on the mobility of the granular bed \cite{Bagnold_1,Raudkivi_1,Bagnold_3}, i.e., bed load affects the fluid flow. This is known as feedback effect, and it is at the origin of an intricate problem because the fluid flow entrains the mobile granular layer, which in turn alters the fluid flow. In this paper, the feedback is defined as the effect caused solely by bed load on the fluid flow.

The mechanism of feedback was first proposed in the case of air flows by Bagnold \cite{Bagnold_2,Bagnold_1,Bagnold_3}, who showed that the transport of grains as bed load significantly changes the air velocity profiles. For bed load in air flows, distinctly from water flows, the ratio of specific masses $\rho_s/\rho$ is large (around $2000$), implying that, in general, the fluid flow has not enough momentum to directly dislodge the grains. The grains are mainly ejected from the bed by shocks with the falling ones and effectuate ballistic flights whose length-scale is many times greater than the characteristic grain diameter. During the ballistic flight, they get momentum from the gas flow, and afterwards, when they fall and reach the fixed part of the granular bed, this momentum is transferred again: one part is responsible for the ejection of other grains; another part is dissipated by the shocks; and, if the grains rebound upwards, another part remains with the same grains. Bagnold \cite{Bagnold_1} interpreted the momentum transfer to salting grains as a drag caused on the fluid by the moving grains, which changes the wind velocity profile near the fixed part of the granular bed, but also above it. More recent experimental works on the feedback effect in air corroborated the work of Bagnold \cite{Rasmunsen_1,Bauer_1,Zhang_1,Yang_1}.

In the case of liquids, the thickness of the mobile granular layer is of a few grain diameters and it is difficult to measure the liquid flow inside the mobile layer. Additionally, the feedback effect in liquids is less pronounced than in gases (the ratio $\rho_s/\rho$ is $1000$ times smaller in liquids), so that it is also difficult to experimentally measure it in regions above the bed.

To our knowledge, the only experimental work that measured the liquid flow inside the mobile granular layer is Charru et al. \cite{Charru_4}, in the case of viscous flows. The fluid velocity profiles were measured inside the mobile layer by PIV (Particle Image Velocimetry) while the grains were identified by Particle Tracking techniques. The authors found parabolic profiles with $\partial^2u/\partial y^2>0$, evidencing the momentum transfer from the fluid to the grains. Whether this parabolic profile is applicable to the fluid inside the moving bed in the case of turbulent flows is still to be verified.

Concerning turbulent liquid flows, there is a lack of experimental works on the feedback effect. The only papers presenting quantitative measurements are perhaps Carbonneau and Bergeron \cite{Carbonneau_Bergeron} and Wang et al. \cite{Wang}.

The experiments of Carbonneau and Bergeron \cite{Carbonneau_Bergeron} were performed in an open channel of variable inclination, in which they imposed different water flow conditions and bed-load transport rates. They employed only one grain type (gravel size particles of $d=7.4mm$) and varied the water flow rate as well as the injection of grains, obtaining  shear velocities in the range $0.031m/s\leq u_*\leq 0.119m/s$, particle Reynolds numbers in the range $229\leq Re_*\leq 880$ (hydraulic rough regime) and bed-load transport rates in the range $0.04kg/s\leq W_b \leq 0.4kg/s$. The instantaneous water velocity profiles were measured over the floor of the channel and over a rough floor (of same granulometry of the employed grains) with an ADV (Acoustic Doppler Velocimetry) device, in the cases where bed load was absent as well as in the cases where it was present. The flow profiles were compared in terms of mean flow and of fluctuations.

Wang et al. \cite{Wang} compared the flows with and without bed load in a direct way. However, they imposed a moving bed over the floor where pure water measurements were previously taken, so that it is probable that the reference height of the flow was different in the two cases. If this is true, a reference height should have been found and the fluid flow profiles should have been vertically shifted before any comparison. The paper does not discuss the correction of the reference height, which may lead to inconsistencies. In their comparisons, some results showed that the mean flow was accelerated in the presence of bed load. In order to explain these results, the authors made an analysis of the exchanges between the mean flow and the fluctuations in the two cases. However, the ADV device was operated at an acquisition frequency of $25Hz$, which seems too low to detect small scale turbulence and to validate any exchange analysis between the mean flow and the fluctuations.

Wang et al. \cite{Wang} investigated the effects of bed load on turbulent, open-channel water flows. They employed two different types of granular beds, composed of pebbles with mean diameters of $d=2mm$ or $d=2.5mm$, and varied the water flow rates, obtaining shear velocities in the range $0.021m/s\leq u_*\leq 0.040m/s$ and particle Reynolds numbers within $41\leq Re_*\leq 100$ (transition and hydraulic rough regime). The flow fields of water over mobile beds, which generated bed forms, were measured by PIV. The paper presents velocity profiles for the mean flow and proposes the use of some correlations for the friction factor when in the presence of bed load. As in the case of Carbonneau and Bergeron \cite{Carbonneau_Bergeron}, it is probable that the granular bed was not the same during the different tests so that they should have found the reference height of each test. This is not discussed in the paper, although the reference height is necessary for the determination of $u_*$ and $B$ from Eq. \ref{mean_velocity}, and then for the friction factor. Finally, Wang et al. \cite{Wang} measured, simultaneously, the effects of both the movement of grains and the shape of bed forms on the mean flow. Therefore, they could not isolate the effects caused solely by bed load on the flow.

\subsection{Paper's objective and organization}

The objective of this paper is to quantify the effect caused solely by bed load on a turbulent liquid flow (without the presence of ripples or dunes), in conditions close to the incipient motion of grains: $\theta\,\sim\,0.01$ \cite{Bagnold_1,Yalin_1,Buffington_1}. The experimental device is a horizontal, rectangular cross-section channel, five meters long, operating with water and having loose and fixed beds at the test section. The turbulent fully-developed velocity profile is measured by Particle Image Velocimetry at the vertical symmetry plane of the channel's test section. The descriptions of the experimental apparatus and procedures are in the next section. The results and the discussions are presented in the following sections. A conclusion section follows.

\section{Experimental set-up}
\label{section_setup}

\subsection{Experimental device}

\begin{figure}
  \begin{center}
    \includegraphics[width=0.90\columnwidth]{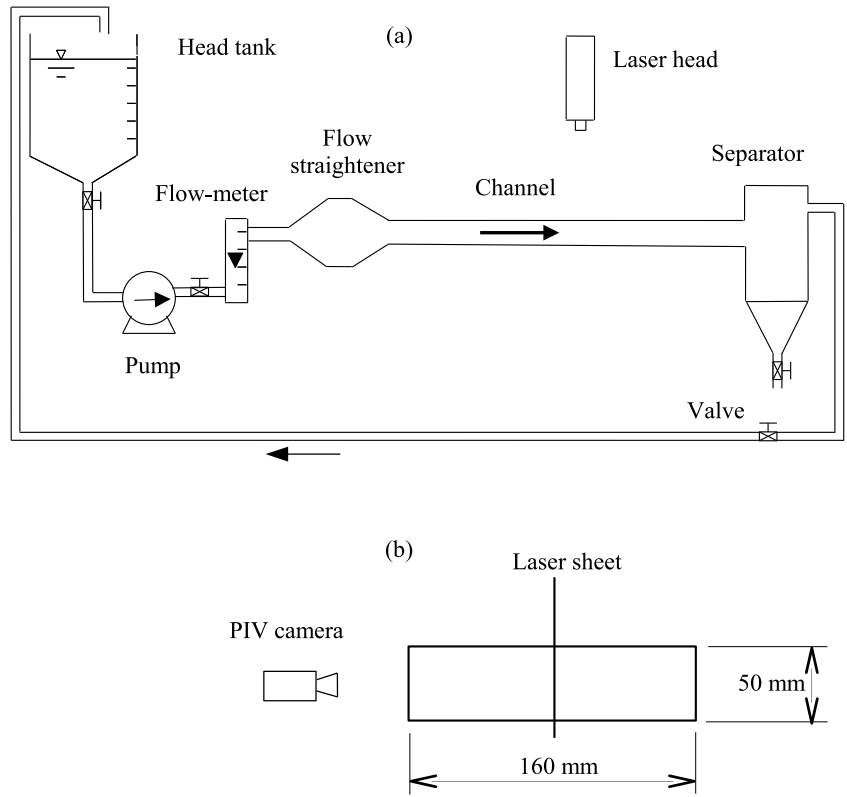}
    \caption{Scheme of the experimental loop: (a) side-view; (b) channel cross section.}
    \label{fig:loop}
  \end{center}
\end{figure}

The experimental device is schematically presented in Fig. \ref{fig:loop}. It consists of a transparent channel, $5m$ long, of rectangular cross section ($160 mm$ wide by $50mm$ high). At the channel's inlet there is a flow straightener composed of a divergent-convergent nozzle, filled with $d=3mm$ glass spheres, whose function is to homogenize the flow profile at the inlet. The test section starts at $40$ hydraulic diameters ($3m$) downstream of the inlet and is $1m$ long. Access windows ensure the introduction of grains. There is another one meter long section connecting the test section exit to the solids separator.

The grains consist of glass spheres exhibiting specific masses of $\rho_s\,=\,2500kg/m^3$ and are classified in two populations. Each population had the minimum and the maximum sizes limited by the use of sieves with distinct meshes: one population had its size ranging from $d = 300 \mu m$ to $d = 425 \mu m$, and it is assumed that the mean diameter was $d_{50}=363 \mu m$; the other population had its size ranging from  $d = 106 \mu m$ to $d = 212 \mu m$, and it is assumed that $d_{50}=159 \mu m$. At the channel's developing section a fixed granular bed was inserted, composed of glass spheres glued on the surface of PVC plates, whose average thickness was $7mm$. For the tests with fixed beds, the plates were introduced in the test section, covering then the entire bottom of the channel. For the tests with mobile beds, the plates in the test section were removed, being replaced by loose grains of same granulometry. This assured a $3m$ ($40$ hydraulic diameters) entrance length of same granulometry and of same thickness as the loose bed.

The tests were performed at ambient conditions, i.e., atmospheric pressure of $1 atm$ and temperature of approximately $25^oC$. The water flowed in a closed loop driven by a pump from the reservoir, through the channel and the grains separator, and back to the reservoir. In cases where a loose bed was present, the grains were entrained by the water stream as bed load. The water flow rate was controlled by changing the excitation frequency of the pump and was measured with an electromagnetic flow-meter accurate to within $0.5\%$ of the measured value. The tests nominal flow rates were of $5$, $5.5$, $6$, $6.5$ and $7 m^3/h$, corresponding to cross-section mean velocities in the range $0.19m/s\leq\overline{U}\leq 0.29m/s$ and to Reynolds number $Re=\overline{U}2H_{gap}/\nu$ in the range $16000 < Re < 27000$, where $H_{gap}$ is the distance from the granular bed to the top wall. The flow rates' lower and upper bounds were fixed to, respectively, correspond to the bed-load threshold \cite{Bagnold_1,Yalin_1,Buffington_1} and to avoid the formation of ripples on the granular bed within the time scale of the tests \cite{Franklin_4,Franklin_5,Franklin_8,Franklin_6}.

Two types of tests were performed. The first type employed the fixed plates throughout the channel length, and are used as the reference state for the water flow without bed load. These tests are identified for short as ``Fix''. The other type concerns the tests in which a loose bed replaced the plates (fixed beds) only at the test section. The plates and the loose bed had the same thickness and were composed of grains of the same granulometry. For each grain population, three series of loose bed tests were performed. The tests performed with the $d_{50}=363 \mu m$ grains were identified as series $A$, $B$ and $C$ while the ones performed with the $d_{50}=159 \mu m$ grains were identified as series $D$, $E$ and $F$.

Prior to each test, the loose granular bed was smoothed and leveled, with the channel already filled up with water. The access to the loose bed was through windows located at the channel's top wall. Next, the test flow rates were established by adjusting the pump's frequency. The pump's frequency controller was pre-programmed, achieving the desired flow rate within a $5s$ period. This mitigated any transient effect as far as the bed-load phenomenon is concerned.

\subsection{Measurement device}

\begin{figure}
  \begin{center}
    \includegraphics[width=0.70\columnwidth]{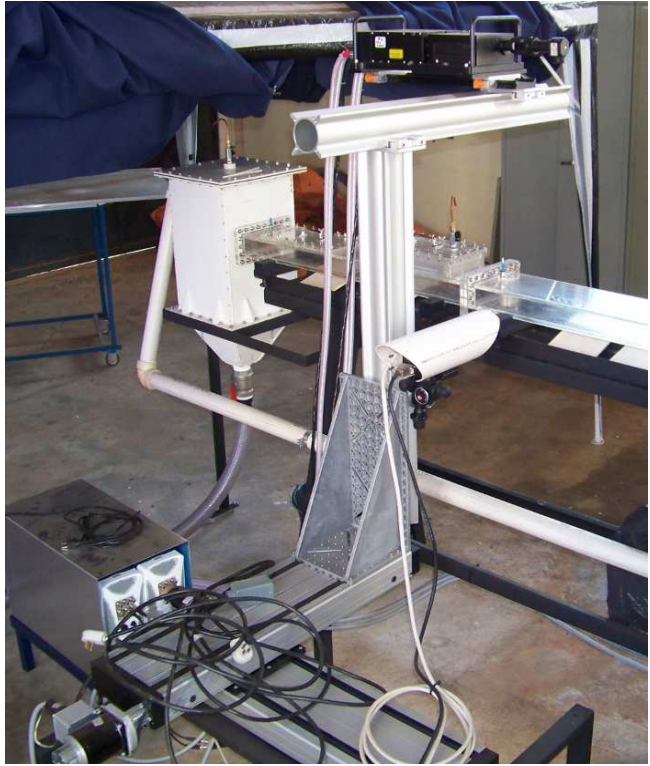}
    \caption{Photograph of the employed PIV device.}
    \label{fig:piv_foto}
  \end{center}
\end{figure}

Particle Image Velocimetry was employed to obtain the instantaneous velocity fields of the water stream. Figure \ref{fig:piv_foto} presents a photograph of the employed PIV device and Fig. \ref{fig:loop} presents a layout of the PIV apparatus. The employed light source was a dual cavity Nd:YAG Q-Switched laser, capable to emit at $2\,\times\,130mJ$ at a $15Hz$ pulse rate. The twin laser beams were controlled by the DaVis software and by a synchronizer with 10 ns time resolution. The laser beam power was fixed within $65\%$ and $75\%$ of the maximum power in order to assure a good balance between image contrasts and undesirable reflection from the granular bed as well as from the channel walls. As seeding particles, we employed the suspension of particulate already present in tap (city) water, together with hollow glass spheres of $10\mu m$ and specific gravity $S.G.=1.05$.

The PIV images were captured by a CCD (charge coupled device) camera acquiring pairs of frames at $4Hz$ with a spatial resolution of $2048px\,\times\,2048px$. The field size employed was of $80mm\,\times\,80mm$ for the tests with fixed beds and of $70mm\,\times\,70mm$ for the tests with mobile beds (note that along the vertical direction the channel height corresponds only to $50mm$). The employed interrogation area was $8px\,\times\,8px$, corresponding then to $256$ interrogation areas of $0.28mm\,\times\,0.28mm$. The computations were made with a $50\%$ overlap, which increases the number of interrogation areas to $512$, and corresponds to a resolution of $0.14mm\,\times\,0.14mm$. The time interval between a pair of frames was set to correspond to a most probable displacement of particles around $5px$.

Each experimental run acquired $500$ pairs of images. The instantaneous velocity fields and the time-averaged velocity field were obtained, respectively, by the image cross correlation and by the average operators built in the software DaVis. MatLab scripts were written to further process the averaged velocity in order to get the friction velocity, the roughness length, the constant $B$ and the origin of the distance from the bed.

\subsection{Origin of the distance from the bed}

The instantaneous velocity fields are accurately measured but it is still necessary to define an origin for the vertical distance $y$ from the bed in order to build a spatio-temporal averaged velocity profile \cite{Raupach,Nikora,Coleman_4}. Difficulties for the definition of $y=0$ (reference height) arise for two reasons: (i) the rough nature of the granular bed, which has different heights at different locations; and (ii) the light scattered by the glass beads composing the bed.

In the case of a turbulent boundary layer over a fixed bed the flow profiles are given by Eq. \ref{mean_velocity}. The reference height can then be found by shifting vertically the mean spatio-temporal velocity profile, in order to fit it with Eq. \ref{mean_velocity} with the corresponding values of $B=5.5$ for the hydraulic smooth regime or $y_0=d_{50}/30$ for the hydraulic rough regime \cite{Schlichting_1,Jimenez_1}. We proceeded in this manner for the fixed bed case. In doing this, the vertical position of the velocity vector closer to the bed was determined. This position, $y=D_{piv}$, was considered as the distance from the bed from which the employed PIV device was able to measure flow velocities.

In the case of mobile granular beds, the velocity profiles are unknown and the definition of the reference height cannot be done by performing a logarithmic fit with Eq. \ref{mean_velocity}. However, a reference height must be found to build a spatio-temporal averaged velocity profile. In order to find a reference height, we assumed that the position $y=D_{piv}$ must be the same for the tests with fixed and loose beds. The reasons for that are: (i) the mobile and the fixed beds were composed of the same grains, with the loose bed still flat and having the same height of the fixed bed; (ii) the tests were performed close to bed-load inception; (iii) the employed laser power was always the same; and (iv) the interrogation area and the position of the laser sheet were the same for all the tests with grains of same granulometry, and total field was roughly the same.

Based on this assumption, the mean spatio-temporal profiles were shifted vertically on the tests for loose beds in order to have the first measured vector at $y=D_{piv}$. The reference height was then at $D_{piv}$ below the lower vector, and any fitting of the measured profiles was based on this reference.

\subsection{Evaluation of the roughness length, of the constant B and of the shear velocity}
\label{sub_B}

Equation \ref{mean_velocity} can be rewritten as

\begin{linenomath*}
\begin{equation}
u\,=\,\frac{u_*}{\kappa}ln(y)\,-\,\frac{u_*}{\kappa}ln\left( y_0 \right)
\label{mean_velocity_mod}
\end{equation}
\end{linenomath*}

\noindent where, for a given fluid flow, we can identify the constants $A=u_*/\kappa$ and $C=(u_*/\kappa )\,ln(y_0)$. When plotted in the log-normal scales, the inclination of Eq. \ref{mean_velocity_mod} is proportional to $u_*$ and the intersection between Eq. \ref{mean_velocity_mod} and the line $u=0$ gives $y_0$. For each water flow condition, the experimental data was fitted to Eq. \ref{mean_velocity_mod} in the region $70<y^+<200$ and the shear velocity and the roughness length were found.

The constant $B$ is obtained directly from the right part of Eq. \ref{mean_velocity}

\begin{linenomath*}
\begin{equation}
B\,=\,-\,\frac{1}{\kappa}ln\left( y_0^+ \right)
\label{eq:B}
\end{equation}
\end{linenomath*}

\section{Results}
\label{section_results}

PIV measurements of the flow field over fixed and mobile granular beds of same granulometry were performed. The averaged velocity profiles over the fixed and the mobile beds, for roughly the same water flow rates, were compared in order to obtain the perturbation caused by the granular mobility. 

\subsection{The fixed bed}
\begin{figure}
  \begin{center}
    \includegraphics[width=1.0\columnwidth]{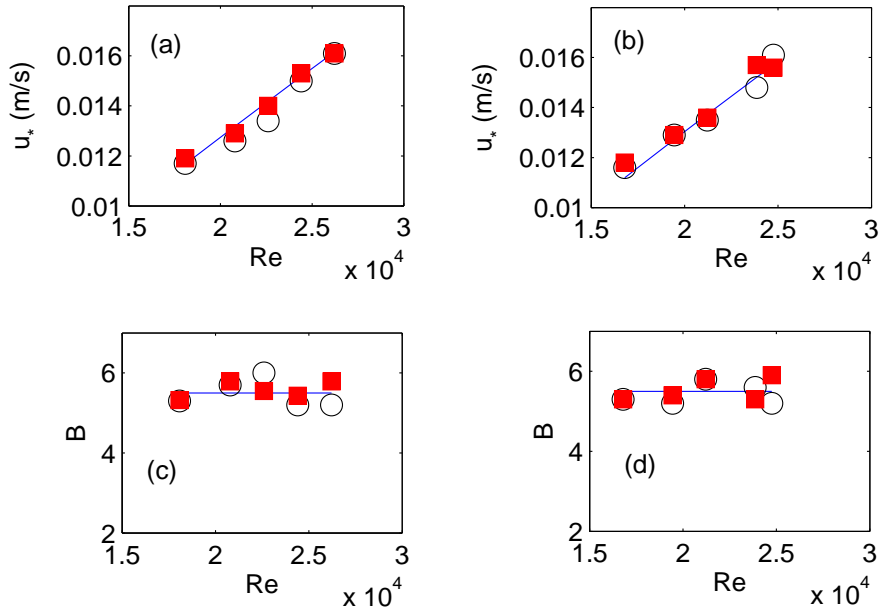}
    \caption{$u_*$ and $B$ as functions of $Re$ on the bottom (granular beds) and top walls. (a) and (c) $d_{50}=363 \mu m$; (b) and (d) $d_{50}=159 \mu m$.}
    \label{fig:u_B_fixo_sup}
  \end{center}
\end{figure}

The experimental data was fitted according to Eq. \ref{mean_velocity_mod} and the values of $u_*$ and $B$ were obtained iteratively. The computed values of $Re_*$ were between $2$ and $6$, indicating that for all the flow rates the boundary layer was in the hydraulic smooth regime, or in the beginning of the transition \cite{Schlichting_1}. Given this regime, the values of $u_*$ and the origin ($y\,=\,0$) were obtained by fitting the profiles in order to have $B$ as close as possible to $5.5$. The obtained values of $u_*$ and $B$ are shown in Fig. \ref{fig:u_B_fixo_sup} as well as in Tabs. \ref{tableau_seuil_1} and \ref{tableau_seuil_2}.

Figures \ref{fig:u_B_fixo_sup}(a) and \ref{fig:u_B_fixo_sup}(b) present the shear velocity $u_*$ and Figs. \ref{fig:u_B_fixo_sup}(c) and \ref{fig:u_B_fixo_sup}(d) the constant $B$ as functions of the Reynolds number $Re$. Figures \ref{fig:u_B_fixo_sup}(a) and \ref{fig:u_B_fixo_sup}(c) concern the $d_{50}=363 \mu m$ bed while Figs. \ref{fig:u_B_fixo_sup}(b) and \ref{fig:u_B_fixo_sup}(d) concern the $d_{50}=159 \mu m$ bed. In these figures, the open circles correspond to the flow on the granular bed, the filled squares correspond to the flow on the top wall and the continuous lines in Figs. \ref{fig:u_B_fixo_sup}(a) and \ref{fig:u_B_fixo_sup}(b) correspond to $u_*$ obtained from the Blasius correlation for smooth walls while in Figs. \ref{fig:u_B_fixo_sup}(c) and \ref{fig:u_B_fixo_sup}(d) they correspond to $B=5.5$. The maximum deviations of $u_*$ from the Blasius correlation and of $B$ from $5.5$ are of $6\%$ and $9\%$, respectively.

\begin{figure}
  \begin{center}
    \includegraphics[width=0.9\columnwidth]{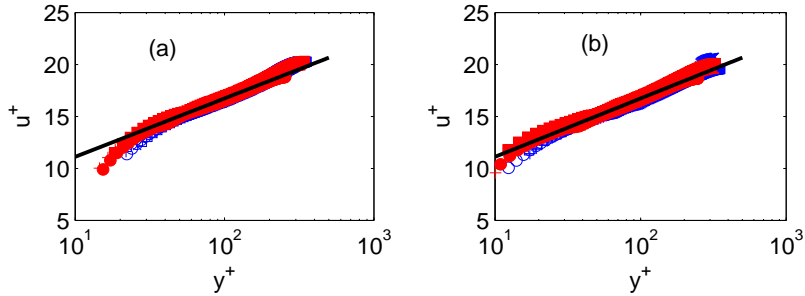}
    \caption{Profiles of the mean velocities on the bottom (granular beds) and top walls, normalized by the internal scales. (a) $d_{50}=363 \mu m$; (b) $d_{50}=159 \mu m$.}
    \label{fig:paredes}
  \end{center}
\end{figure}

Tables \ref{tableau_seuil_1} and \ref{tableau_seuil_2} present a summary of the estimations for the $d_{50}=363 \mu m$ and $d_{50}=159 \mu m$ beds, respectively. In these tables, ``Fix'' stands for fixed bed and ``mob A'', ``mob B'', ``mob C'', ``mob D'', ``mob E'' and ``mob F'' stand for, respectively, the series of tests with mobile beds, called here A, B, C, D, E and F. The column ``symbol'' lists the symbols employed in Figs. \ref{fig:paredes} and \ref{fig:perfil_log_all}, $Q$ is the water flow rate, $\overline{U}$ is the cross-section mean velocity of the water, $U_{max}$ is the maximum water velocity of each profile, $Re_*$ is the particle Reynolds number (defined by Eq. \ref{reynolds}) and $\theta$ is the Shields number (defined by Eq. \ref{shields}). The estimations of $u_*$ and $B$ for the loose bed cases are described in the next subsection.

Figure \ref{fig:paredes} presents the log-normal profiles of the mean velocities on the bottom (granular beds) and top walls, normalized by the internal scales (shear velocity and viscous length). The abscissa, in logarithmic scale, represents the normalized distance from the walls (bottom or top wall). Figure \ref{fig:paredes}(a) corresponds to the $d_{50}=363 \mu m$ bed, Fig. \ref{fig:paredes}(b) corresponds to the $d_{50}=159 \mu m$ bed and the list of symbols is presented in Tabs. \ref{tableau_seuil_1} and \ref{tableau_seuil_2} (the filled symbols correspond to the profiles on the top wall). The continuous line, shown as a reference, corresponds to Eq. \ref{mean_velocity}. Figure \ref{fig:paredes} shows that the normalized profiles on the bottom and top walls are almost superposed, although different granulometries were present. This is a characteristic of the smooth regime.

\begin{figure}
  \begin{center}
    \includegraphics[width=1.0\columnwidth]{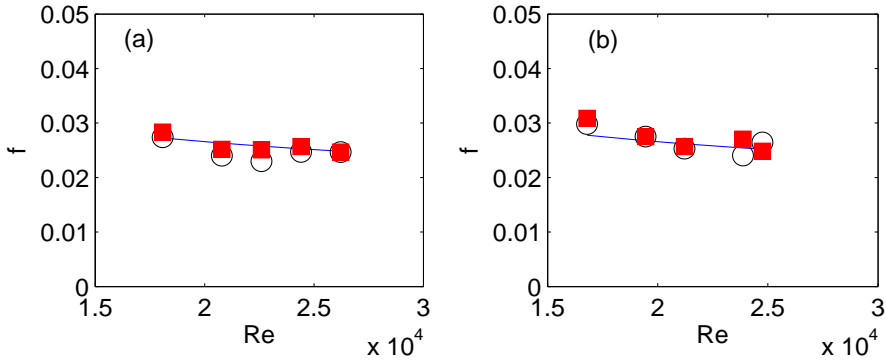}
    \caption{$f$ as a function of $Re$ on the bottom (granular beds) and top walls. (a) $d_{50}=363 \mu m$; (b) $d_{50}=159 \mu m$.}
    \label{fig:fat}
  \end{center}
\end{figure}

Figure \ref{fig:fat} presents the Darcy friction factor $f=8(u_*^2)/(\overline{U}^2)$ as a function of the channel Reynolds number $Re$. In this figure, the open circles correspond to the friction factor on the granular bed, the filled squares correspond to the friction factor on the top wall and the continuous line corresponds to Blasius correlation for smooth walls $f=0.316Re^{-1/4}$. Figures \ref{fig:fat}(a) and \ref{fig:fat}(b) correspond, respectively, to the $d_{50}=363 \mu m$ and $d_{50}=159 \mu m$ beds. They show that the Blasius correlation provides a good fit to the measured friction velocities, corroborating that the measured flows are in the smooth regime.

\subsection{The loose bed}

\begin{figure}
  \begin{center}
    \includegraphics[width=0.90\columnwidth]{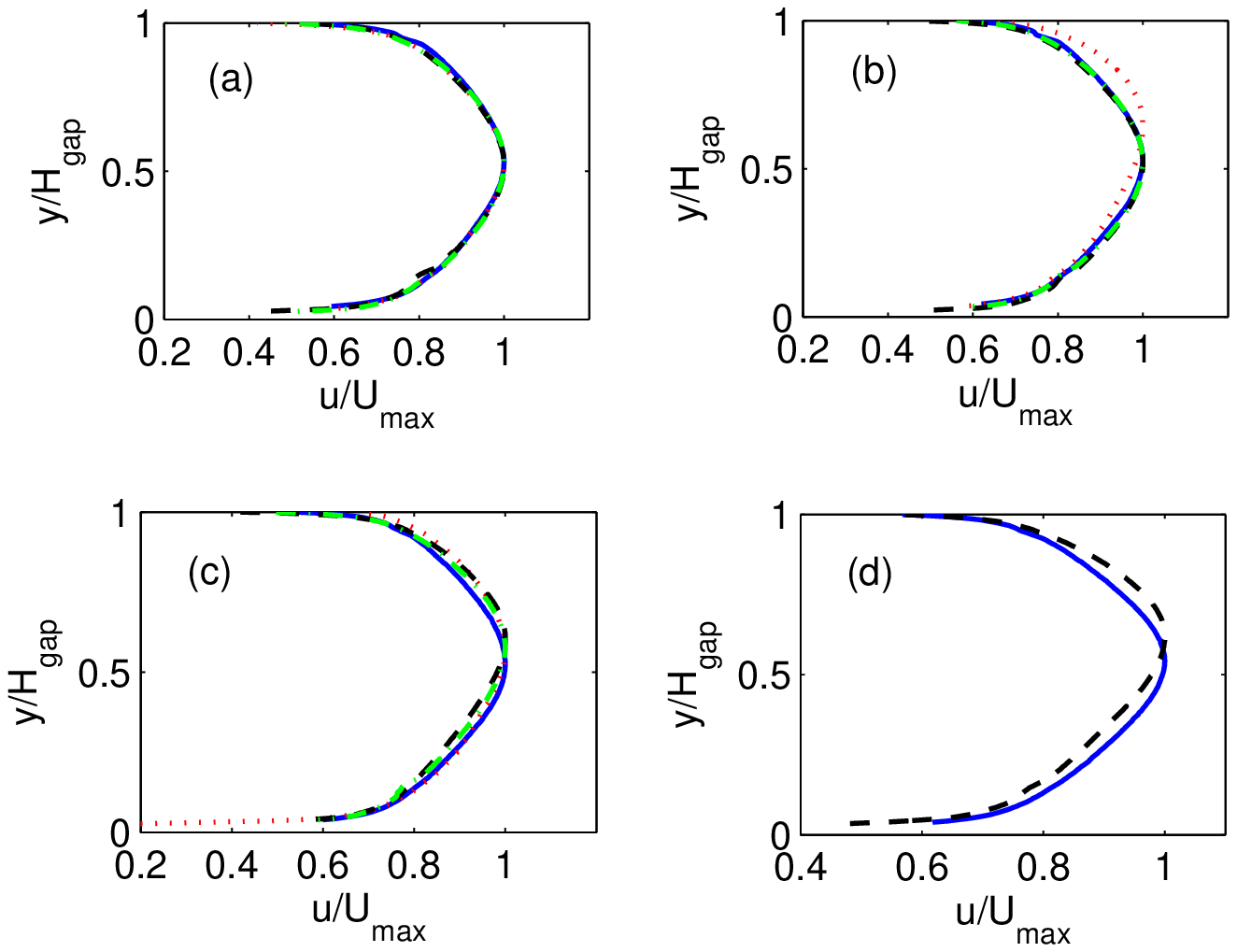}
    \caption{Profiles of the mean velocities over fixed and mobile granular beds of cases A, B and C ($d_{50}=363 \mu m$). (a) $\overline{U}=0.21m/s$; (b) $\overline{U}=0.25m/s$; (c) $\overline{U}=0.27m/s$; (d) $\overline{U}=0.29m/s$.}
    \label{fig:perfil_linear_300}
  \end{center}
\end{figure}

Figure \ref{fig:perfil_linear_300} presents the profiles of the mean velocities over fixed and mobile granular beds of cases A, B and C ($d_{50}=363 \mu m$), roughly subjected to the same water flow rates. The continuous lines correspond to the fixed bed case, the dotted lines to series A, the dashed lines to series B and the dashed-dotted lines to series C (cf. Tab. \ref{tableau_seuil_1}), and the scales are linear. $U_{max}$ is the maximum water velocity of each profile and $H_{gap}$ is the distance from the granular bed to the top wall. Figures \ref{fig:perfil_linear_300}(a), \ref{fig:perfil_linear_300}(b), \ref{fig:perfil_linear_300}(c) and \ref{fig:perfil_linear_300}(d) correspond, respectively, to cross-section mean velocities $\overline{U}$ of $0.21m/s$, $0.25m/s$, $0.27m/s$ and $0.29m/s$ (or water flow rates of roughly $5m^3/h$, $6m^3/h$, $6.5m^3/h$ and $7m^3/h$, respectively).

In our experiments, the bed-load threshold was close to $\overline{U}=0.21m/s$. At this condition, it is expected that the density of mobile grains, i.e. the ratio of the quantities of mobile to static grains, will be low, causing then a very small effect on the water stream. This is what the measurements indicate, as can be observed from Fig. \ref{fig:perfil_linear_300}(a). In this figure, the profiles are almost superposed due to the little effect of bed load on the water stream. In fact, they are not perfectly superposed because the flow rate was slightly different in the fixed bed case (cf. Tab. \ref{tableau_seuil_1}). We note here that the kinks seen in $y/H_{gap}\,\approx\,0.14$ and in $y/H_{gap}\,\approx\,0.95$ for some of the profiles come from the saturation of the PIV images in this region caused by undesired reflection.

The density of mobile grains is expected to increase with the cross-section mean velocity of the water, increasing then the feedback effect. This is shown in Figs. \ref{fig:perfil_linear_300}(b), \ref{fig:perfil_linear_300}(c) and \ref{fig:perfil_linear_300}(d). In these figures, some of the profiles over the mobile beds have larger asymmetries, displacing the maximum velocity loci upwards, which indicates higher flow resistances. The asymmetry is more pronounced for stronger flow rates.

Based on air profiles over loose beds \cite{Bagnold_2,Bagnold_1,Bagnold_3,Rasmunsen_1,Bauer_1}, we may expect the existence of a logarithmic sublayer. If the logarithmic region of the measured water profiles can be identified, then the shear velocity $u_*$, the constant $B$ and the roughness length $y_0$ can be estimated. Tables \ref{tableau_seuil_1} and \ref{tableau_seuil_2} present a summary of these estimations for the $d_{50}=363 \mu m$ and $d_{50}=159 \mu m$ beds, respectively.

\begin{figure}
  \begin{center}
    \includegraphics[width=0.95\columnwidth]{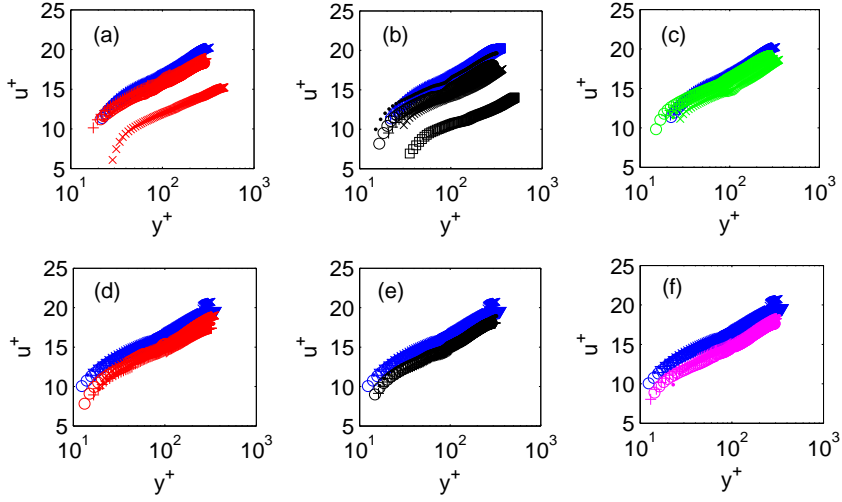}
    \caption{Profiles of the mean velocities over fixed and mobile granular beds, normalized by the internal scales. (a), (b), (c), (d), (e) and (f) correspond to series A, B, C, D, E and F, respectively.}
    \label{fig:perfil_log_all}
  \end{center}
\end{figure}

Figure \ref{fig:perfil_log_all} presents the profiles of the mean velocities over fixed and mobile granular beds, submitted to the same water flow rates, in the traditional log-normal scales. Figures \ref{fig:perfil_log_all}(a), \ref{fig:perfil_log_all}(b) and \ref{fig:perfil_log_all}(c) correspond to cases A, B and C ($d_{50}=363 \mu m$) and Figs. \ref{fig:perfil_log_all}(d), \ref{fig:perfil_log_all}(e) and \ref{fig:perfil_log_all}(f) correspond to cases D, E and F ($d_{50}=159 \mu m$), respectively. In these figures, the mean velocity is normalized by the shear velocity and the vertical distance is normalized by the viscous length. The employed symbols are listed in Tabs. \ref{tableau_seuil_1} and \ref{tableau_seuil_2}.

Comparing the profiles over the mobile beds with the ones over a fixed bed, we observe that they are similar: they present a logarithmic region above the bed. However, the profiles over mobile beds are displaced to higher locations (the same velocities are reached at higher transversal coordinates), and this displacement tends to increase as the granular mobility is increased. In addition, although of same granulometry, we can observe differences between the series A, B and C and between series D, E and F. These differences can be imputed to the bed armouring, as explained in the following.

The agreement of the profiles shown in Fig. \ref{fig:perfil_log_all} with Eq. \ref{mean_velocity} in a region close to the granular bed indicates that the flow is in local equilibrium in this region. The same occurs close to the top wall, not shown here, and the interaction between the two near-wall regions occurs at the central region. It is interesting to note that a similar behavior was experimentally measured in a different case by Hanjalic and Launder \cite{Hanjalic_Launder}, who were interested in the effects of asymmetry in wall roughness on closed-conduit flows. They employed an asymmetric two-dimensional channel, which was composed of fixed walls with different bottom and top wall roughness, and the tests were all performed in the hydraulic rough regime. The similarities between the profiles obtained by Hanjalic and Launder \cite{Hanjalic_Launder} and the ones obtained in the present work seem to indicate that one effect of bed load on a closed-conduit flow is to cause an asymmetry that would be equivalent to an increase in the roughness, even if the flow is in the hydraulic smooth regime. We analyze next only the region near the granular bed.

\begin{table}[htbp]
\begin{center}
\begin{tabular}{|c|c|c|c|c|c|c|c|c|c|}
	\hline
	case & $Q$ & $\overline{U}$ & Symbol & $B$ & $y_0$ & $u_*$ & $U_{max}$ & $Re_*$ & $\theta$\\
	$\cdots$ & $m^3/h$ & $m/s$ & $\cdots$ & $\cdots$ & $\cdots$ & $m/s$ & $m/s$ & $\cdots$ & $\cdots$\\
	\hline
	Fix & $5.0$ & $0.20$ & $\circ$ & $5.3$ & $d/37$ & $0.0117$ & $0.22$ & $4$ & $0.03$\\
	\hline
	Fix & $5.6$ & $0.23$ & $+$ & $5.7$ & $d/47$ & $0.0126$ & $0.25$ & $5$ & $0.03$\\
	\hline
	Fix & $6.1$ & $0.25$ & $\cdot$ & $6.0$ & $d/57$ & $0.0134$ & $0.28$ & $5$ & $0.03$\\
	\hline
	Fix & $6.8$ & $0.27$ & $\times$ & $5.2$ & $d/47$ & $0.0150$ & $0.30$ & $5$ & $0.04$\\
	\hline
	Fix & $7.3$ & $0.29$ & $\square$ & $5.2$ & $d/50$ & $0.0161$ & $0.33$ & $6$ & $0.05$\\
	\hline
	mob A & $5.3$ & $0.21$ & $\circ$ & $4.4$ & $d/28$ & $0.0130$ & $0.24$ & $5$ & $0.03$\\
	\hline
	mob A & $5.8$ & $0.23$ & $+$ & $5.0$ & $d/38$ & $0.0136$ & $0.26$ & $5$ & $0.03$\\
	\hline
	mob A & $6.1$ & $0.25$ & $\cdot$ & $4.1$ & $d/29$ & $0.0145$ & $0.27$ & $5$ & $0.04$\\
	\hline
	mob A & $6.8$ & $0.27$ & $\times$ & $2.9$ & $d/21$ & $0.0171$ & $0.30$ & $6$ & $0.05$\\
	\hline
	mob B & $5.3$ & $0.21$ & $\circ$ & $3.9$ & $d/24$ & $0.0133$ & $0.24$ & $5$ & $0.03$\\
	\hline
	mob B & $5.9$ & $0.24$ & $+$ & $3.4$ & $d/22$ & $0.0149$ & $0.26$ & $5$ & $0.04$\\
	\hline
	mob B & $6.3$ & $0.25$ & $\cdot$ & $5.3$ & $d/46$ & $0.0145$ & $0.29$ & $5$ & $0.04$\\
	\hline
	mob B & $6.7$ & $0.27$ & $\times$ & $3.0$ & $d/21$ & $0.0168$ & $0.30$ & $6$ & $0.05$\\
	\hline
	mob B & $7.3$ & $0.29$ & $\square$ & $-0.5$ & $d/7$ & $0.0228$ & $0.33$ & $8$ & $0.10$\\
	\hline
	mob C & $5.4$ & $0.21$ & $\circ$ & $5.1$ & $d/37$ & $0.0126$ & $0.24$ & $5$ & $0.03$\\
	\hline
	mob C & $5.9$ & $0.23$ & $+$ & $5.0$ & $d/38$ & $0.0136$ & $0.26$ & $5$ & $0.03$\\
	\hline
	mob C & $6.4$ & $0.26$ & $\cdot$ & $4.3$ & $d/32$ & $0.0153$ & $0.29$ & $6$ & $0.04$\\
	\hline
	mob C & $6.8$ & $0.27$ & $\times$ & $3.8$ & $d/27$ & $0.0162$ & $0.31$ & $6$ & $0.05$\\
	\hline
\end{tabular}
\caption{Computed shear velocity $u_*$, roughness length $y_0$ and constant $B$, for the $d_{50}=363 \mu m$ beds.}
\label{tableau_seuil_1}
\end{center}
\end{table}

\begin{table}[htbp]
\begin{center}
\begin{tabular}{|c|c|c|c|c|c|c|c|c|c|}
	\hline
	case & $Q$ & $\overline{U}$ & Symbol & $B$ & $y_0$ & $u_*$ & $U_{max}$ & $Re_*$ & $\theta$\\
	$\cdots$ & $m^3/h$ & $m/s$ & $\cdots$ & $\cdots$ & $\cdots$ & $m/s$ & $m/s$ & $\cdots$ & $\cdots$\\
	\hline
	Fix & $4.9$ & $0.19$ & $\circ$ & $5.3$ & $d/16$ & $0.0116$ & $0.22$ & $2$ & $0.06$\\
	\hline
	Fix & $5.4$ & $0.22$ & $+$ & $5.2$ & $d/18$ & $0.0129$ & $0.25$ & $2$ & $0.07$\\
	\hline
	Fix & $5.9$ & $0.24$ & $\cdot$ & $5.8$ & $d/23$ & $0.0135$ & $0.27$ & $2$ & $0.08$\\
	\hline
	Fix & $6.7$ & $0.27$ & $\times$ & $5.6$ & $d/24$ & $0.0148$ & $0.31$ & $2$ & $0.09$\\
	\hline
	Fix & $7.0$ & $0.28$ & $\square$ & $5.2$ & $d/22$ & $0.0161$ & $0.31$ & $3$ & $0.11$\\
	\hline
	mob D & $5.2$ & $0.20$ & $\circ$ & $3.9$ & $d/10$ & $0.0129$ & $0.24$ & $2$ & $0.07$\\
	\hline
	mob D & $5.8$ & $0.23$ & $+$ & $3.3$ & $d/9$ & $0.0148$ & $0.26$ & $2$ & $0.09$\\
	\hline
	mob D & $6.3$ & $0.25$ & $\cdot$ & $3.5$ & $d/10$ & $0.0156$ & $0.28$ & $2$ & $0.10$\\
	\hline
	mob D & $6.9$ & $0.27$ & $\times$ & $4.2$ & $d/14$ & $0.0161$ & $0.31$ & $3$ & $0.11$\\
	\hline
	mob D & $7.5$ & $0.29$ & $\square$ & $4.4$ & $d/17$ & $0.0170$ & $0.33$ & $3$ & $0.12$\\
	\hline
	mob E & $5.3$ & $0.21$ & $\circ$ & $3.8$ & $d/10$ & $0.0131$ & $0.24$ & $2$ & $0.07$\\
	\hline
	mob E & $5.8$ & $0.23$ & $+$ & $3.6$ & $d/10$ & $0.0144$ & $0.26$ & $2$ & $0.09$\\
	\hline
	mob E & $6.3$ & $0.25$ & $\cdot$ & $4.4$ & $d/14$ & $0.0148$ & $0.28$ & $2$ & $0.09$\\
	\hline
	mob F & $5.3$ & $0.21$ & $\circ$ & $3.4$ & $d/8$ & $0.0134$ & $0.24$ & $2$ & $0.08$\\
	\hline
	mob F & $5.7$ & $0.23$ & $+$ & $3.9$ & $d/11$ & $0.0140$ & $0.26$ & $2$ & $0.08$\\
	\hline
	mob F & $6.3$ & $0.25$ & $\cdot$ & $3.5$ & $d/10$ & $0.0151$ & $0.28$ & $2$ & $0.10$\\
	\hline
\end{tabular}
\caption{Computed shear velocity $u_*$, roughness length $y_0$ and constant $B$, for the $d_{50}=159 \mu m$ beds.}
\label{tableau_seuil_2}
\end{center}
\end{table}

For flows over mobile beds, the values of $B$ and $u_*$ are expected to be different, and to vary with the mobility of the grains. This is confirmed by the displacement of the turbulent boundary layer when bed load is present. However, we note that for some tests the values of $B$ and $u_*$ are almost unchanged with respect to the fixed bed case. In some cases, this can be explained by inception conditions for bed load during the test \cite{Yalin_2,Buffington_1,Charru_1,Cao}. For instance, some of the $d_{50}=363 \mu m$ tests had $\theta\,\leq\,0.04$. For the remaining of the tests, the water flow was increased, or the granulometry changed to $d_{50}=159 \mu m$, and the values of the Shield parameter were higher, so that the reason for only small variations in $B$ and $u_*$ shall be different. One possible explanation can be the bed armouring.

\begin{figure}
  \begin{center}
    \includegraphics[width=0.90\columnwidth]{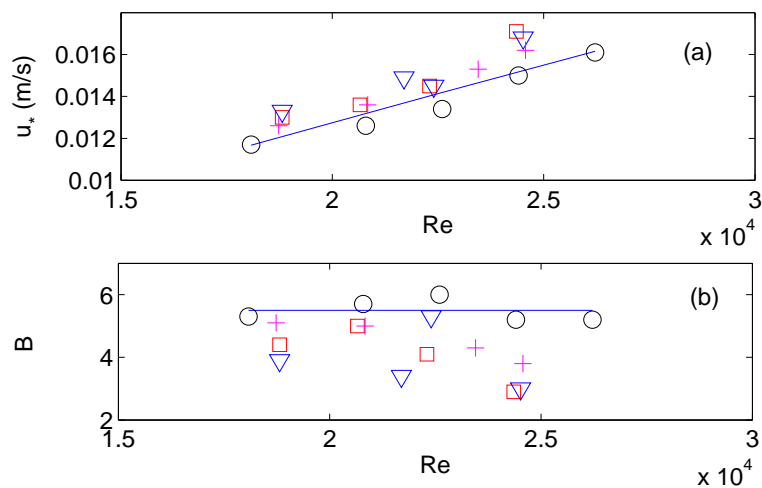}
    \caption{(a) $u_*$ and (b) $B$ as functions of $Re$ for $d_{50}=363 \mu m$. $\bigcirc$, $\Box$, $\triangledown$ and $+$ correspond, respectively, to the fixed bed and to the series A, B and C. The continuous lines correspond to the Blasius correlation in (a) and to $B=5.5$ in (b).}
    \label{fig:u_B_360}
  \end{center}
\end{figure}

Figure \ref{fig:u_B_360} presents the variations of the shear velocity $u_*$ and of the constant $B$ with the Reynolds number $Re$, for the $d_{50}=363 \mu m$ beds. Squares, triangles and crosses correspond, respectively, to the series A, B and C and the circles correspond to the fixed bed case. The continuous lines correspond to the Blasius correlation in Fig. \ref{fig:u_B_360}(a) and to $B=5.5$ in Fig. \ref{fig:u_B_360}(b). We decided to neglect the values of $u_*$ and $B$ obtained for $\overline{U}=0.29m/s$, series B, because a discrepancy seems to exist. Experiments with the $d_{50}=363 \mu m$ beds show a tendency to increase $u_*$ and to decrease $B$ with the increase in the cross-section mean velocity of the water. In conditions near the bed-load threshold ($\overline{U} <0.26m/s\,\Rightarrow\,Re<21000$), the tests present values of $u_*$ and $B$ close to the fixed bed case. As the water flow rate is increased, the values go far from the fixed bed case, indicating that within the duration of the tests the density of mobile grains increased with the water mean velocities and that the feedback effect was present.

\begin{figure}
  \begin{center}
    \includegraphics[width=0.90\columnwidth]{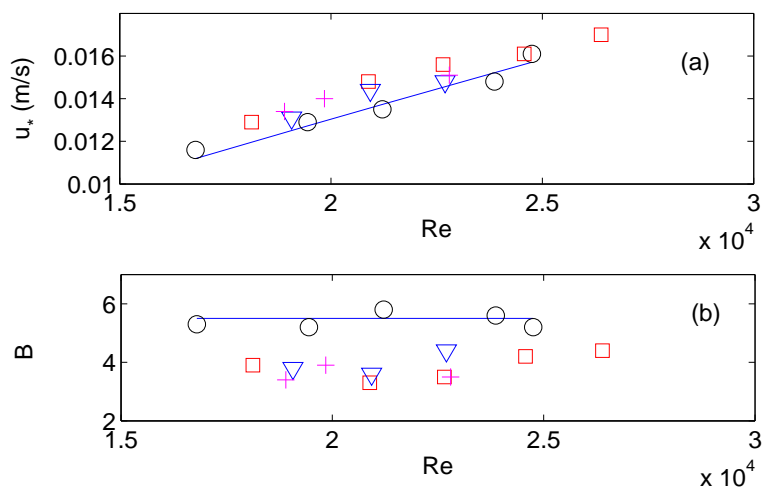}
    \caption{(a) $u_*$ and (b) $B$ as functions of $Re$ for $d_{50}=159 \mu m$. $\bigcirc$, $\Box$, $\triangledown$ and $+$ correspond, respectively, to the fixed bed and to the series D, E and F. The continuous lines correspond to the Blasius correlation in (a) and to $B=5.5$ in (b).}
    \label{fig:u_B_160}
  \end{center}
\end{figure}

Figure \ref{fig:u_B_160} presents the variations of the shear velocity $u_*$ and of the constant $B$ with the Reynolds number $Re$, for the $d_{50}=159 \mu m$ beds. Squares, triangles and crosses correspond, respectively, to the series D, E and F and the circles correspond to the fixed bed case. The continuous lines correspond to the Blasius correlation in Fig. \ref{fig:u_B_160}(a) and to $B=5.5$ in Fig. \ref{fig:u_B_160}(b). Experiments with the $d_{50}=159 \mu m$ beds were performed farther from the bed-load threshold ($\theta>0.06$) and the weakest cross-section mean velocities (that correspond to $Re<21000$) present values of $u_*$ and $B$ different from the fixed bed case. However, their behavior with $Re$ is different from that of the $d_{50}=363 \mu m$ beds: the values of $u_*$ and $B$ tend to the fixed bed values as the water flow rate is increased. This would mean that the density of mobile grains, and then the feedback effect, decreased with the cross-section mean velocity of the water. This was not expected and an explanation for this can be the bed armouring \cite{Charru_1}. In our tests, each series of experiments was performed by increasing sequentially the water flow rate, and it is probable that by the time the PIV measurements were made for the highest flow rates (highest water velocities), the granular mobility decreased due to armouring effects. The $d_{50}=159 \mu m$ bed is more susceptible to armouring because the density of hollows on the surface is greater than on the $d_{50}=363 \mu m$ bed, so that the probability of trapping mobile grains is higher. Therefore, bed armouring is suspected for the $d_{50}=159 \mu m$ cases within the duration of the tests.

Finally, we note that the variations of $u_*$ and $B$ are relatively small, so that their uncertainties must be taken into account. If the uncertainties are estimated as the deviations of $u_*$ from the Blasius correlation and of $B$ from the law of the wall, both for the fixed bed case, then they are of $\pm 6\%$ and $\pm 9\%$ ($\pm 0.5$), respectively. 

Concerning $u_*$ for the mobile beds, both the deviations from the fixed bed cases and the variations within each series are higher than the estimated uncertainties. Based on this, we consider that the pointed effects of the granular mobility on $u_*$ are valid.

Concerning the values of $B$ for the mobile beds, two remarks can be made. The first one is that the deviations of $B$ from the fixed bed cases are greater than the estimated uncertainties, so that the values of $B$ are indeed lower in the mobile cases. The second one is that within each $d_{50}=363 \mu m$ series the differences between the maxima and the minima of $B$ are not much greater than the estimated uncertainties. For the $d_{50}=159 \mu m$ beds, they are equal. This implies that, although the variations of $B$ with the fluid velocity corroborate that of $u_*$, they are close to the estimated uncertainties. In summary, the values of $B$ are lower in the mobile case, but no definite conclusion can be made for the variations of $B$ with the fluid velocities.

\section{Discussion}
\label{section_discussion}

Bagnold \cite{Bagnold_1} showed, in the case of hydraulic rough gas flows, that the influence of bed load on the mean flow can be regarded as the generation of an effective roughness which is greater than the one expected for a fixed granular bed with same granulometry. He proposed that the increase of the effective roughness with the granular mobility is due to the momentum transfer from the fluid to the mobile grains, and from the latter to the fixed part of the bed, where it is dissipated (mainly by shocks, in the case of gases).

In the case of water, the higher specific mass of the fluid corresponds to lower velocity reductions for the same momentum transfer. This means that, given a loose granular bed, the bed-load threshold occurs at lower velocity gradients (and lower cross-section mean velocities), so that it can happen in the hydraulic smooth regime. Although in our experiments the boundary layer was in the hydraulic smooth regime, Fig. \ref{fig:perfil_log_all} shows a displacement of the normalized profiles that increases with the mobility of the grains, very similar to the aeolian case (hydraulic rough).

For liquid flows in the hydraulic smooth regime, we propose that the feedback mechanism is still similar to the one proposed by Bagnold \cite{Bagnold_1}: the momentum transfer from the fluid to the mobile grains decelerates the liquid in the region near the bed, displacing the water flow profiles but keeping the same normalized form. This displacement increases with granular mobility (and with the mean water velocities) due to an increase in the momentum transfer from the fluid to the mobile grains, and then to dissipative mechanisms. In the case of liquids, a great part of the initial energy is dissipated by the grains dislodgement, by friction and by the drainage of the interstitial fluid, which generates small scale vortices. The displacement of the velocity profile would be the equivalent of an increase in the effective roughness of a rough regime, although the water flow is in the hydraulic smooth case (cf. Tabs. \ref{tableau_seuil_1} and \ref{tableau_seuil_2}). The effect on the water flow of this momentum transfer is a strong slow down in the region close to the granular bed. Well above the granular bed, in the core flow, the water flow is less affected by it (if compared to a flow over a fixed bed of same granulometry). This implies that in this region the mean velocity profiles shall be similar to the ones over a fixed bed.

We propose here that the liquid flow profile over a mobile bed consists of: (a) a lower region ($y^+<40$), close to the granular bed, where the effects of momentum transfer are very strong and slow down the liquid flow; (b) an upper region ($y^+>200$), far enough from the granular bed (in the core flow) so that the mean fluid flow shall have a profile similar to the one over a fixed bed of same granulometry; and (c) a matching region ($50<y^+<150$) which, as in the case of turbulent boundary layers over fixed granular beds, shall have a logarithmic profile. These three regions can be seen in Fig. \ref{fig:perfil_log_all}.

In the case of closed-conduit flows, the flow in the upper region (core flow) changes slightly, different from open flows: the slow down in the lower region implies, by mass conservation, an acceleration in the upper region and an increase in the asymmetry. If the fluid flow is near the bed-load threshold, the momentum transfer in the lower region is relatively small, so that the deceleration near the bed and the acceleration in the core flow are barely noticed, and the shear velocity is almost the same as in the case of a fixed bed. As the fluid flow rate is increased, the density of mobile grains increases and there is an increase in the momentum transfer near the bed: the slow down near the bed and, consequently, the acceleration in the core flow increase the shear velocity in both walls (lower and upper). Also, the slow down in the lower region displaces the normalized logarithmic layer to higher transversal positions, corresponding to the lower values measured for the $B$ constant over the bed. In this work we analyzed only the effects in the region near the granular bed, so that we didn't report the variations of $u_*$ on the top wall.

\section{Conclusion}

This paper was devoted to the feedback effect on a turbulent liquid flow. We presented an experimental study on the perturbation of a turbulent boundary layer of a liquid by a mobile granular bed, in conditions close to incipient motion, in a closed-conduit flow. An experimental loop was built with water as the liquid phase and glass beads as the granular media composing the bed, which was either static or mobile. The velocity field of the water stream was measured by Particle Image Velocimetry in the cases of static and mobile granular beds, and the changes caused by the bed-load transport could be determined.

We successfully measured the perturbation caused on the turbulent stream by the bed load. The good agreement between the measured profiles and the law of the wall in the region close to the granular bed indicates that the flow is in local equilibrium in this region. The same occurs close to the top wall and the interaction between the two near-wall regions occurs at the central region. We analyzed in this paper only the region close to the granular bed.

Our measurements indicate the existence of at least three distinct regions in the mean water flow: one region near the granular bed, where the flow is strongly slowed down due to momentum transfer, called here lower region; a region far from the granular bed, in the core flow, where the perturbation effects are weaker, called here upper region; and an intermediate region, which must match the upper and the lower ones, and then must have a logarithmic profile, called here matching region. It was shown that, although in the hydraulic smooth regime, the bed-load effect on the mean water stream is the vertical displacement of the normalized velocity profiles. This effect would be the equivalent of the roughness length increase in a rough regime.

\begin{acknowledgements}
The authors are grateful to Petrobras S.A. for the financial support (contract number 0050.0045763.08.4), including the scholarship granted to Fab\'iola Tocchini de Figueiredo. The authors thank Guilherme Augusto Ayek for the help with the experimental device and the PIV set up. Erick de Moraes Franklin is grateful to FAPESP (grant 2012/19562-6).
\end{acknowledgements}


\bibliography{referencias}

\begin{thebibliography}{10}
\providecommand{\url}[1]{{#1}}
\providecommand{\urlprefix}{URL }
\expandafter\ifx\csname urlstyle\endcsname\relax
  \providecommand{\doi}[1]{DOI \discretionary{}{}{}#1}\else
  \providecommand{\doi}{DOI \discretionary{}{}{}\begingroup
  \urlstyle{rm}\Url}\fi

\bibitem{Bagnold_1}
R.A. Bagnold, \emph{The physics of blown sand and desert dunes} (Chapman and
  Hall, 1941)

\bibitem{Raudkivi_1}
A.J. Raudkivi, \emph{Loose boundary hydraulics}, 1st edn. (Pergamon Press,
  1976)

\bibitem{Yalin_1}
M.S. Yalin, \emph{Mechanics of sediment transport}, 1st edn. (Pergamon Press,
  1977)

\bibitem{Gomez_1}
B.~Gomez, M.~Church, Water Resour. Res. \textbf{25}(6), 1161 (1989)

\bibitem{Nakato_1}
T.~Nakato, J. Hydraul. Eng. \textbf{116}(3), 362 (1990)

\bibitem{Martin_1}
Y.~Martin, Geomorphology \textbf{53}, 75 (2003)

\bibitem{Schlichting_1}
H.~Schlichting, \emph{Boundary-layer theory} (Springer, 2000)

\bibitem{Tennekes_Lumley}
H.~Tennekes, J.L. Lumley, \emph{First course in turbulence} (The MIT Press,
  1972)

\bibitem{Bagnold_3}
R.A. Bagnold, Philos. Trans. R. Soc. Lond. Ser.A \textbf{249}, 235 (1956)

\bibitem{Bagnold_2}
R.A. Bagnold, Proc. R. Soc. Lond. \textbf{157}, 594 (1936)

\bibitem{Rasmunsen_1}
K.R. Rasmunsen, J.D. Iversen, P.~Rautahemio, Geomorphology \textbf{17}, 19
  (1996)

\bibitem{Bauer_1}
B.O. Bauer, C.A. Houser, W.G. Nickling, Geomorphology \textbf{59}, 81 (2004)

\bibitem{Zhang_1}
W.~Zhang, Y.~Wang, S.J. Lee, Geomorphology \textbf{88}, 109 (2007)

\bibitem{Yang_1}
P.~Yang, Z.~Dong, G.~Qian, W.~Luo, H.~Wang, Geomorphology \textbf{89}, 320
  (2007)

\bibitem{Charru_4}
F.~Charru, H.~Mouilleron-Arnould, O.~Eiff, J. Fluid Mech. \textbf{629}, 229
  (2009)

\bibitem{Carbonneau_Bergeron}
P.E. Carbonneau, N.E. Bergeron, Geomorphology \textbf{35}, 267 (2000)

\bibitem{Wang}
X.~Wang, Q.~Yang, W.~Lu, X.~Wang, Water Resour. Manage. \textbf{25}, 2781
  (2011)

\bibitem{Buffington_1}
J.M. Buffington, D.R. Montgomery, Water Resour. Res. \textbf{33}, 1993 (1997)

\bibitem{Franklin_4}
E.M. Franklin, J. Braz. Soc. Mech. Sci. Eng. \textbf{32}(4), 460 (2010)

\bibitem{Franklin_5}
E.M. Franklin, J. Braz. Soc. Mech. Sci. Eng. \textbf{33}(3), 265 (2011)

\bibitem{Franklin_8}
E.M. Franklin, J. Braz. Soc. Mech. Sci. Eng. \textbf{34}(1), 1 (2012)

\bibitem{Franklin_6}
E.M. Franklin, Appl. Math. Model. \textbf{36}, 1057 (2012)

\bibitem{Raupach}
M.R. Raupach, R.A. Antonia, S.~Rajagopalan, Appl. Mech. Rev. \textbf{44}, 1
  (1991)

\bibitem{Nikora}
V.~Nikora, K.~Koll, I.~McEwan, S.~McLean, A.~Dittrich, J. Hydraul. Eng.
  \textbf{130}, 1036 (2004)

\bibitem{Coleman_4}
S.E. Coleman, V.I. Nikora, S.R. McLean, T.M. Clunie, B.W. Melville, J. Hydraul.
  Eng. \textbf{133}, 121 (2007)

\bibitem{Jimenez_1}
J.~Jim\'enez, Ann. Rev. Fluid Mech. \textbf{36}, 173 (2004)

\bibitem{Hanjalic_Launder}
K.~Hanjalic, B.E. Launder, J. Fluid Mech. \textbf{51}, 301 (1972)

\bibitem{Yalin_2}
M.S. Yalin, E.~Karahan, J. Hydraul. Div \textbf{HY11}, 1433 (1979)

\bibitem{Charru_1}
F.~Charru, H.~Mouilleron-Arnould, O.~Eiff, J. Fluid Mech. \textbf{519}, 55
  (2004)

\bibitem{Cao}
Z.~Cao, G.~Pender, J.~Meng, J. Hydraul. Eng. \textbf{132}, 1097 (2006)

\end{thebibliography}
\bibliographystyle{spphys}

\end{document}